# EVIDENCE OF DAMAGE IN CARBON FIBRE COMPOSITE TILES JOINED TO A METALLIC HEAT SINK UNDER HIGH HEAT FLUX FATIGUE

R. Mitteau, P. Chappuis, L. Moncel and J. Schlosser

Association Euratom - CEA sur la Fusion Contrôlée.
Centre d'Etude de Cadarache
F-13108 Saint Paul Lez Durance CEDEX

## ABSTRACT

The two years experience with Active Metal Casting flat bonds shows that this technology is suitable for the heat fluxes expected in Tore Supra (10 MW/m²). Tests were pursued up to 3330 cycles, with elements still functional. At higher heat fluxes, fatigue damage is observed, but the bond resists remarkably well with no tile detachment. Examination of such deliberately damaged bonds showed distributed cracking, proving the absence of any weak link. The limitations to those higher heat fluxes are more related to the design and the base materials than to the bond itself.

## I.     INTRODUCTION

Continuous operation in tokamaks or stellerators requires the use of actively cooled plasma facing components. One of the main issues in the manufacturing of such components is to join the plasma facing material, often a refractory material with a low thermal expansion coefficient, to a structural metallic heat sink with a much higher thermal expansion coefficient.

Until recently, the technology to join Carbon Fibre Composite (CFC) tiles to the metallic heat sink was direct brazing [1-3]. In 1995 a new technology, Active Metal Casting (AMC), was introduced by Plansee [4]. This technology features a laser-treatment of the joining surface. An Oxygen Free High Conductivity (OFHC) copper layer is cast on the tile using a wettablility agent. The CFC/copper blocks are then brazed or EB-welded to the heat sink.

This technology is today's state of the art, allowing to manufacture large series of elements. It is used for macroblocs a well as for flat tile geometries. At Tore Supra, macroblocs are still investigated [5], but the emphasis was put on flat tile geometries because elements are easier to design, manufacture and control. It is planed for the Toroïdal Pumped Limiter (TPL) as well as for the lateral protections of the RF antennas [6,7]. The CFC is N11 from "Société Européenne de Propulsion" (SEP) and the heat sink material is Chromium Zirconium Copper (CuCrZr).

## II.     HIGH HEAT FLUX TESTS OF FLAT AMC BONDS

Such flat AMC bonds have been extensively tested on the high heat flux facility EB-200 of CEA/Framatome in Le Creusot [8], namely two TPL subscale fingers (PL1 and PL5), four TPL scale-one fingers (C2, R2, PL2-3 and PL2-4) [8] and two ITER baffle mock-ups (SMS 2B and 3B, [9]). More recently, four W7X divertor target prototypes and eight short fingers for the protections of RF antennas with the same technology have been tested.

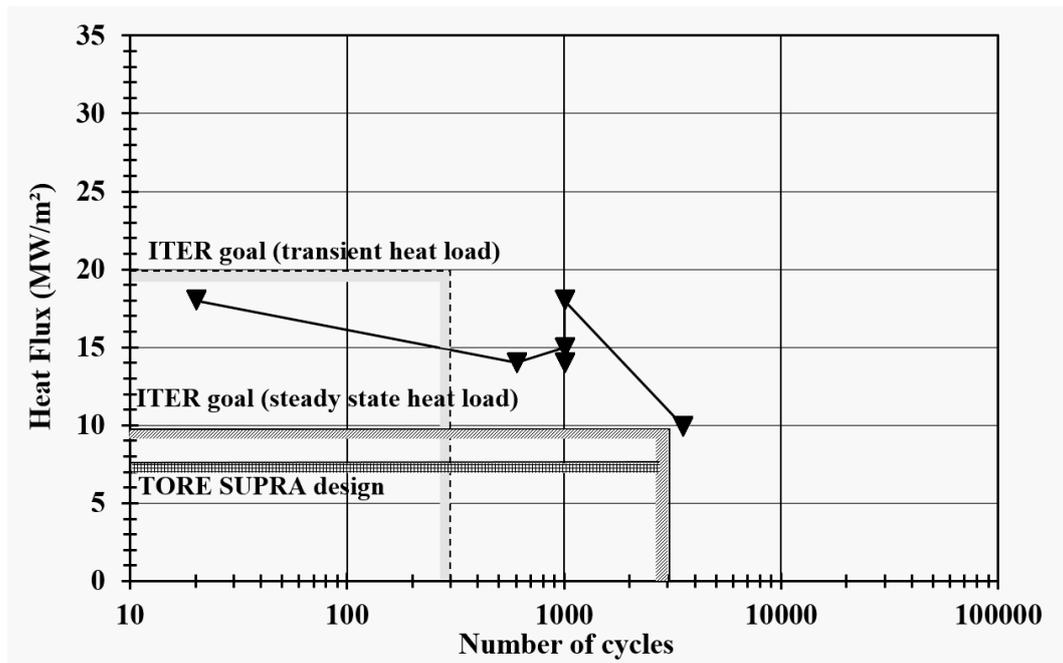

Figure 1 : results of high heat flux tests of CFC armour to Cu heat sink joints.

All these tests represent 242 tiles, from which more than half were cycled. The total tested surface amounts to 1838 cm². Most mock-ups were tested with a flat beam profile above the expected heat flux to explore the limits of the design. All heat loads mentioned below are absorbed power densities deduced from calorimetric measurements.

PL1 and PL5 sustained 1000 cycles at 15 MW/m² without evolution of the surface temperature. R2 sustained 1000 cycles without damage at 14 MW/m², but C2 began to show two evolving tiles after 600 cycles for the same heat flux. Other parts of R2 and C2 were tested at 18 MW/m² and showed clear damages on some tiles after a few cycles. This value of 18 MW/m² appears to be the ultimate heat flux for flat AMC bonds. The other scale one fingers PL2-3 and PL2-4 were tested at 10 MW/m² during 3330 cycles (see section III), the test being stopped with the mock-up still functional.

With NS31 from SEP and Dunlop concept 2, SMS 2B and 3B were cycled during 1000 cycles at 18 MW/m², showing an evolution of the surface temperature, and intense erosion but no tile detachment. Those tests compare well with the literature (figure 1, see also [10]).

The AMC bond in flat tile geometry, although less capable in term of highest heat load, is the second ever tested in term of number of cycles.

## III. REFINED ANALYSIS OF SCALE-ONE FINGERS PL2-3 AND PL2-4

The behaviour of PL2-3 and PL2-4 was investigated more thoroughly because they were submitted to 3330 cycles, which largely exceeds the usual 1000 cycles. 3330 cycles is comparable to the number of full power shots expected for the TPL in Tore Supra. The 10 MW/m² flat heat load was severe compared to design peak values of 6 MW/m² for the flat part and 8 MW/m² for the leading edge.

The surface temperature of both fingers is recorded during the cycling by an infrared camera.

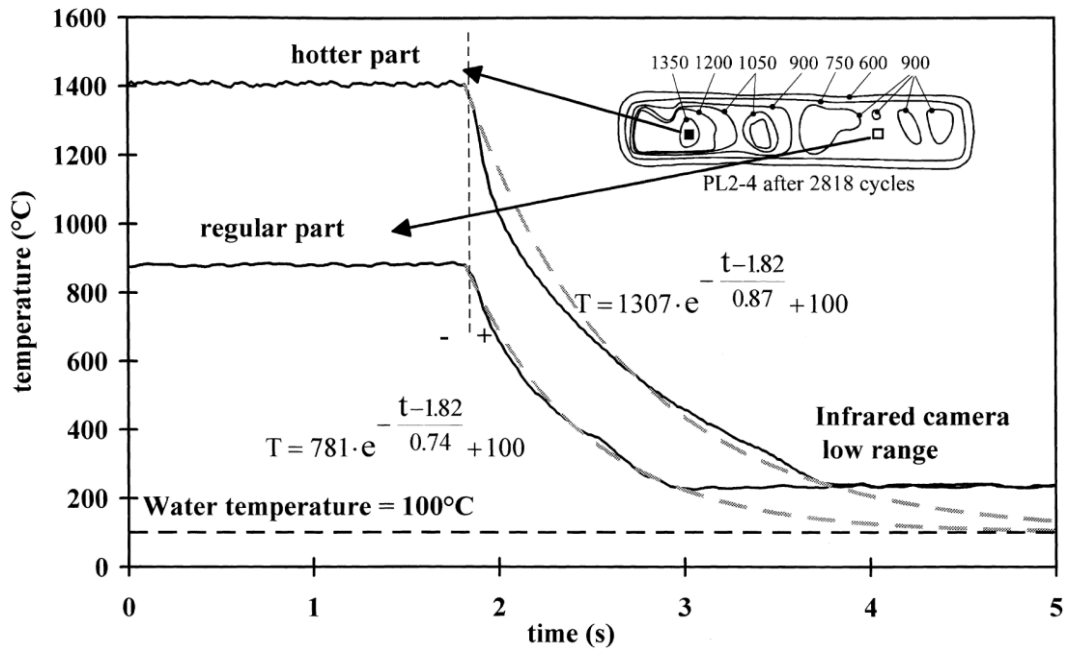

Figure 2 : time evolution of the surface temperature after stopping the incident power after 2818 cycles on PL2-4.

$$T = 1307 \cdot e^{-\frac{t-1.82}{0.87}} + 100$$

$$T = 781 \cdot e^{-\frac{t-1.82}{0.74}} + 100$$

The infrared view shows regular and hotter parts (figure 2). The hotter parts can be attributed to numerous effects such as deteriorated bonds, lower tile conductivity, local differences in the surface emissivity and non evenness of the beam. After the cycling, surface temperature of both parts versus time are extracted by averaging on 10*10 pixels zones (figure 2).

The surface temperature is calculated on the plateau and the time constant during the cooling of the structure. The emissivity of the surface was set to 0.85. Calibration shots before and after the tests showed that the correction coefficient (transmission of the windows, tile emissivity, etc...) didn't noticeably changed during tests. Only the cycles with a heat flux higher than 10 MW/m² (ranging from 10.6 to 11.5 MW/m²) were taken into account in this study. The surface temperatures were normalised to 10 MW/m² according to finite element calculations to allow comparison. The normalised surface temperature of regular and hotter parts of PL2-3 and PL2-4 are given versus the number of cycles (figure 3).

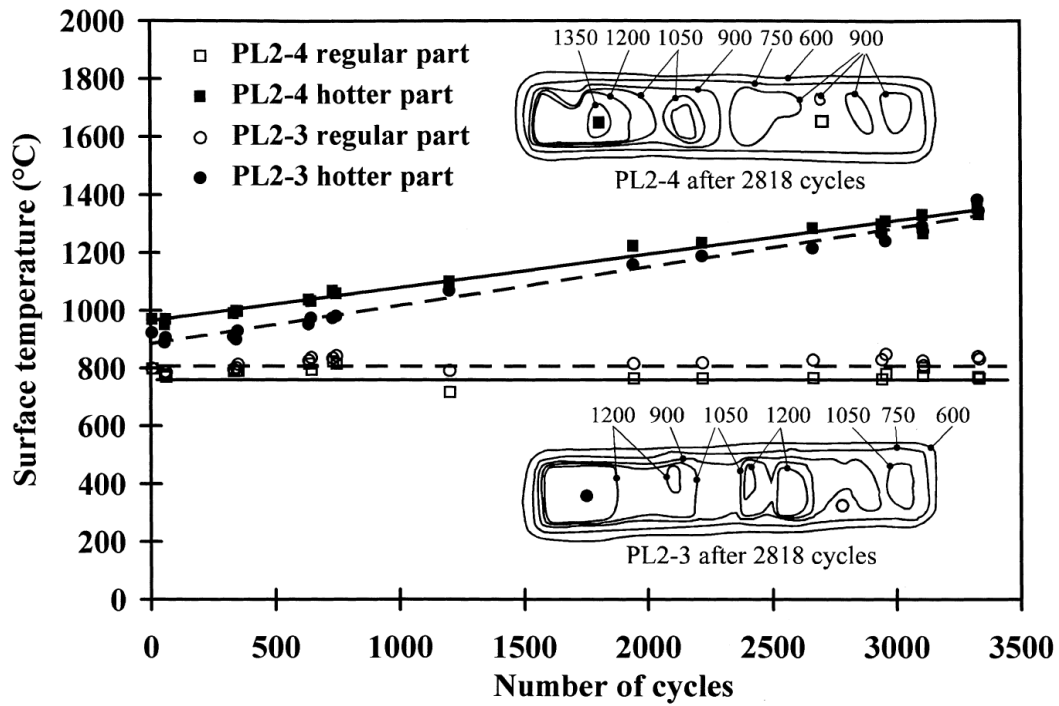

Figure 3 : Evolution of the surface temperature versus number of cycles for TPL prototype elements.

The temperatures of the regular tiles of both fingers are almost constant, around 800°C. Hotter parts on both fingers show a steady increase of the temperature with the cycling, from 950°C to 1350°C. The temperature rise of the hot tiles cannot be attributed to lower tile conductivity or beam profile, which are constant during the test. For this reason, this temperature rise is linked to progressive damage at the interface, either as the progression of a crack or to progressive micro-cracking.

The temperature variation on neighbouring points gives the uncertainty of the method. For 0, 500, 1000 and 3000 cycles where at least two temperatures for close number of cycles are known, the fluctuations can be evaluated to 30 °C. Therefore, the temperature rise of ~ 400°C of the hotter parts of PL2-3 and PL2-4 is significant.

The interface damage is also characterized by a longer time response of the tiles. The time constants of the tiles are calculated during the transients. They are given figure 4 for hotter and regular parts.

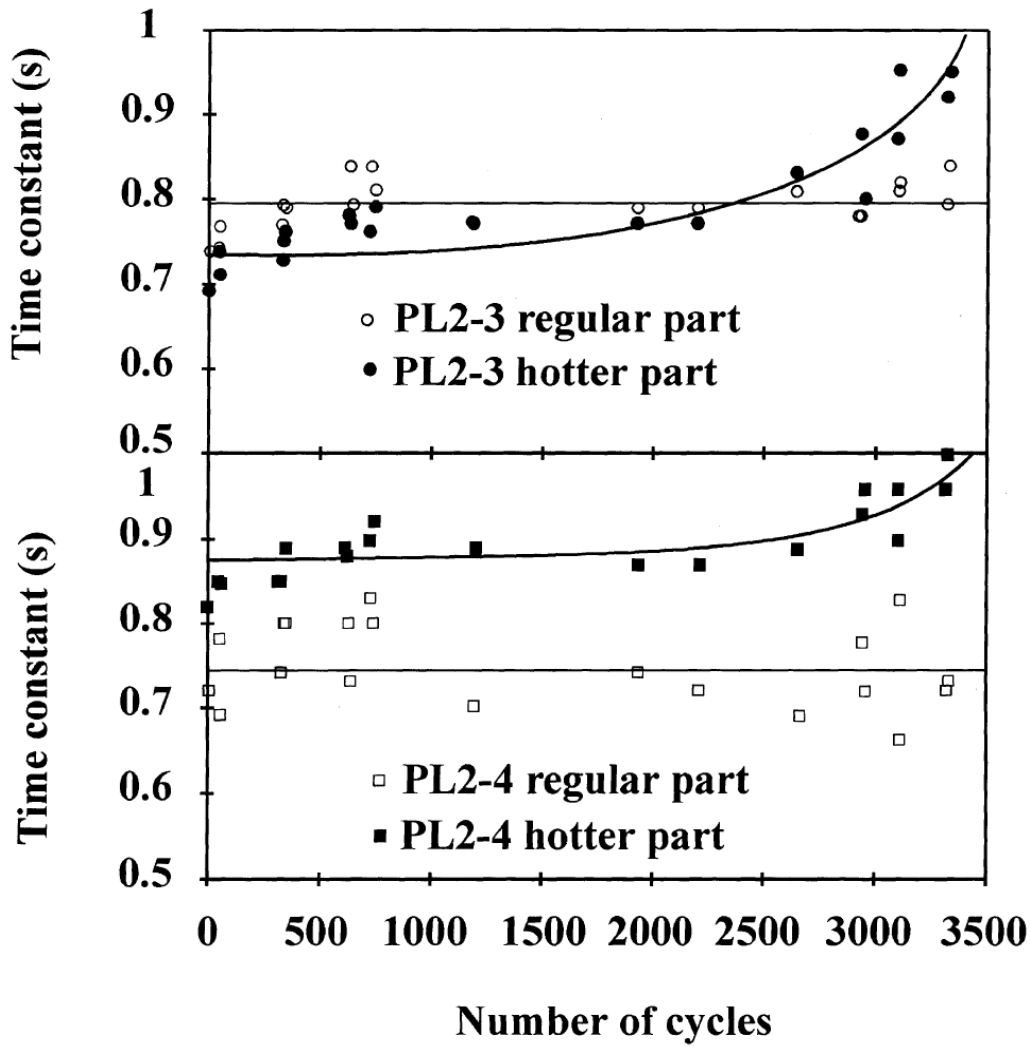

Figure 4 : Evolution of the time constants.

Rather big fluctuations are observed, around 0.07 second while the time constant is around 0.7 second. However, different evolutions between the spots are still significant. The time constants of the regular zones are approximately constant, while those of the hotter parts always increase with the number of cycles. The hotter tiles always finish with longer time constants than the regular tiles. For PL2-3, the increase is around 23 % of the initial value, and for PL2-4 around 12%. The time constants, although less precise than the surface temperature measurements, confirm the degradation of the heat transfer capability of the structure. However, the damaged tiles remain remarkably actively cooled, with time constants smaller than 1 second and a surface temperature still acceptable for operation.

## IV. MACROGRAPHIC OBSERVATIONS

Macro-examinations were done in an attempt to link the surface temperature rise to defects. However, this task is made more difficult because macroscopic defects such as large cracks appear only on tiles that underwent a heat flux higher than ~15 MW/m². Elements fatigued at 10 MW/m² (like PL2-3 and PL2-4) don't show much more defects as uncycled ones.

Examinations show that defects can appear in three areas, namely the composite, the bond and the soft copper layer, the latest being not addressed here

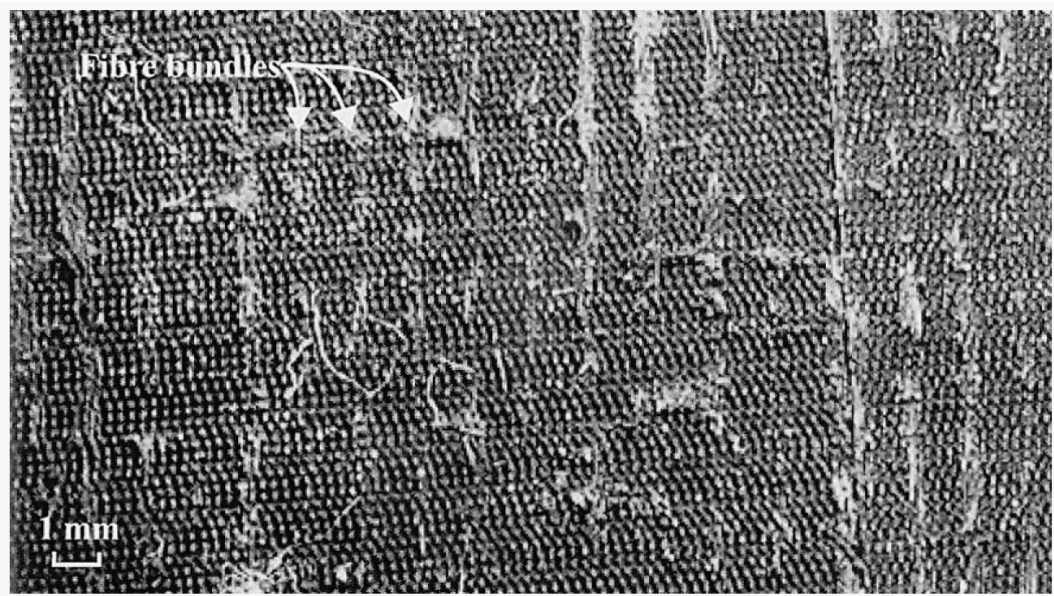

Figure 5 : Macrography of the heat sink after tests at 18 MW/m² (the defective tile was removed in laboratory)

On broken tiles removed from the mock-up after test at 18 MW/m² (figure 5), the copper is mainly black, with many cones visible. The cones are made out of copper that has solidified in the laser - machined holes in the CFC during the AMC process. Fibre bundles often remain attached to the copper. Very few copper is visible on the fractured area of the tile. These elements indicate that such cracks propagate mainly in the CFC or just beneath. Moreover, some black dust falls from the tile, suggesting a disintegration of the CFC material in the vicinity of the bond.

In the CFC (figure 6), cracks are observed above the cones, on top of the cones or between cones. On the edges, cracks start in the CFC in front of the cones. Inside the tiles, the cracks begin at the interface to copper or inside the CFC. Cracks can be observed at the interface between the copper and the CFC. More often such cracks are hybrid, that is they propagate indifferently in the CFC or at the interface of the CFC to copper. In these areas, the cones recede from the CFC, leaving a crack that follows the cones or sometimes goes through the CFC between two cones. On some very damaged tiles, continuous cracks that go straight through the cone and the CFC could be observed.

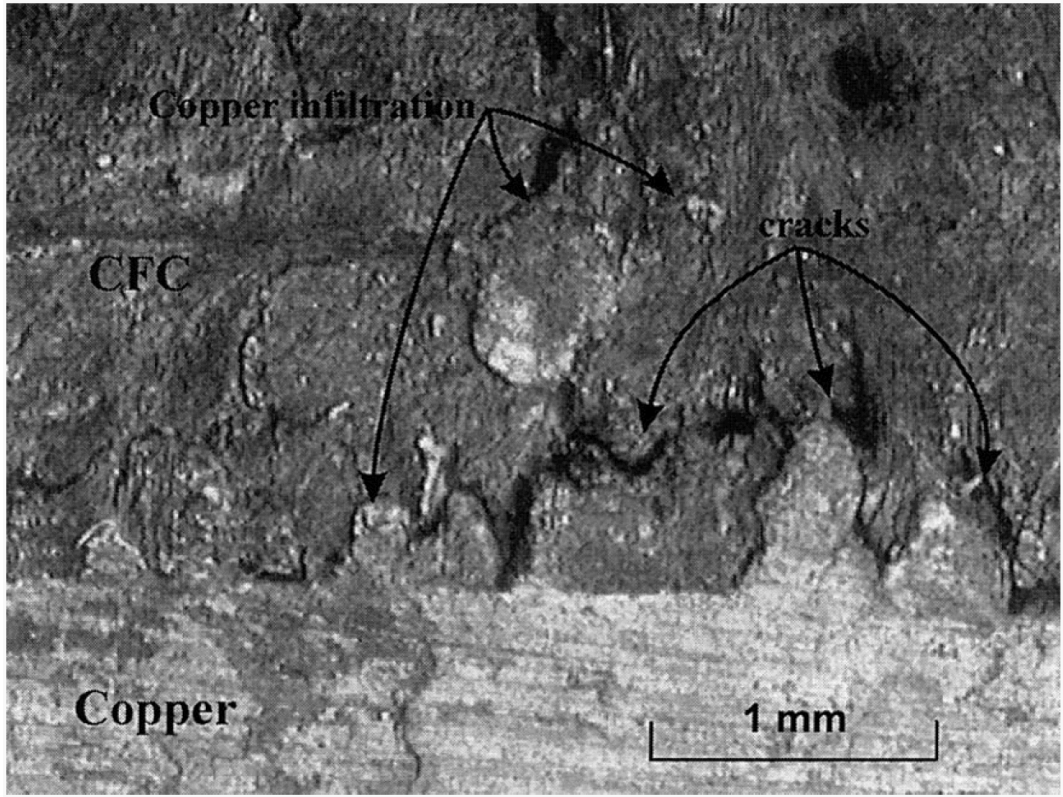

Figure 6 : Macrography of cracks in the CFC and at the interface CFC-copper.

Internal voids in the CFC near the interface are sometime filled with copper or carbides. Such inclusions in the CFC are to expand and shrink during thermal cycling. As the CFC has a lower thermal expansion coefficient, it will undergo cyclic compression. This suggests a damage mechanism which could explain the disintegration of the CFC above the bond. In some cases, such inclusions were observed above cracks, showing that the crack was subsequent to the manufacture of the AMC bond.

The variety of defects suggest that many damage mechanisms can take place. No evidence of any weak link can be found ; damages are distributed throughout the bond. No simple description of the bond fatigue can be given and further investigations are necessary. This is currently being done with specific experiments, like tensile tests, shear tests and dilatometric measurements.

## V. DISCUSSION

Finite element calculations are done to evaluate the stresses in the structure of a finger. A new non-linear model was specially developed and implemented for the CFC [11]. Stresses in a CFC tile of the LPT for 10 MW/m² and a computation of the Tsai-Wu criterion are shown figure 7.

In most of the tile, the criterion exceeds 1, showing the CFC material should break. Even if the criterion may be overestimated, the level of stress explains the observed cracks. It explains also the kind of disintegration of the CFC observed above the bond. However, the cracks also allow the CFC to accommodate somehow the imposed deformation, which could explain that damaged tiles still hold to the heat sink. These results also point out the limitation of the Tsai-Wu criterion, which isn't suitable for this configuration. A new criterion is currently under development at Tore Supra to characterize the stress level.

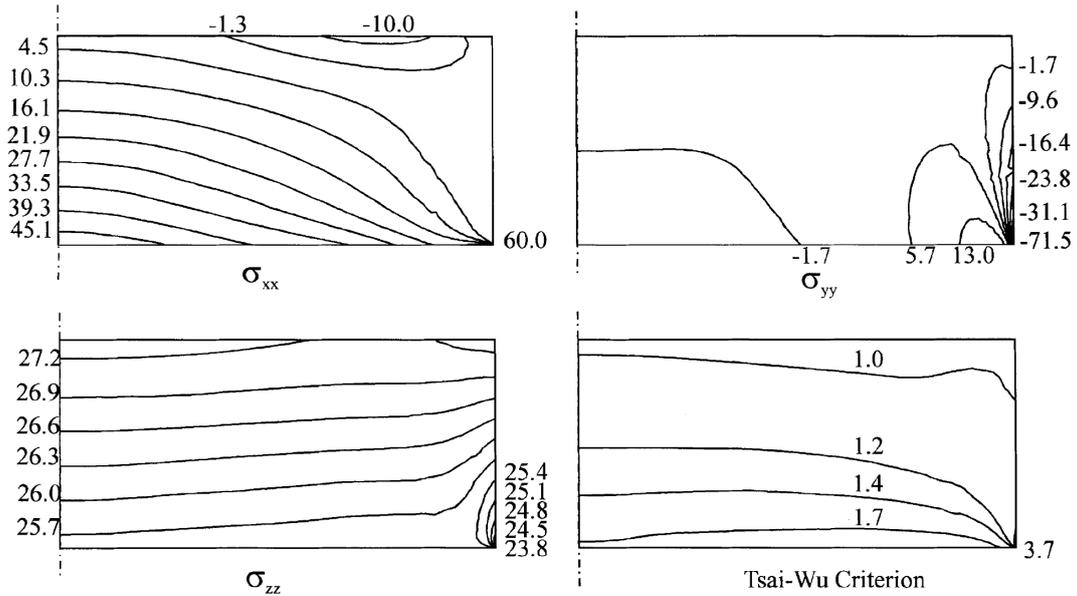

Figure 7 : Stresses (MPa) in a CFC tile under 10 MW/m² (2D generalised plain strain calculation)

The very thin CFC-copper interface is more difficult to analyse and modelize. It was shown in section IV that cracking also occurs there. Seemingly, a key advantage of the AMC technology comes from the important increase of the interface surface created by the cones. Assuming 0.25 mm high cones with a periodicity of 0.25 mm, and a base radius of 0.08 mm, the increase of the area of the interface amounts to 175 % compared to a flat one. Even with an interface weaker than the base materials, the surface increase can compensate its reduction of strength.

## VI. CONCLUSION

AMC flat bonds have proved their ability to sustain high heat fluxes and thermal fatigue. Much higher fluxes than the design value of Tore Supra combined with cycling are necessary to observe damages. This result allows to enter the manufacturing stage, and approximately 1000 fingers totalling roughly 8 m² of bond have been ordered for the toroïdal pumped limiter and the RF antennas lateral protections for Tore Supra. Developments are still under way to prepare future plasma facing components which will have to sustain higher heat fluxes than those of today. The macrobloc geometry could allow higher heat fluxes. Changes in the local geometry near the singular point of flat tiles are also investigated to delay the progression of damage.